\documentclass[journal]{IEEEtran}
\usepackage{cite,graphicx,amsmath,amssymb}
\usepackage{subfigure}
\usepackage{citesort}
\usepackage{fancyhdr}
\usepackage{mdwmath}
\usepackage{mdwtab}
\usepackage{balance}
\usepackage{xcolor}
\usepackage{bm}
\usepackage{amsthm}
\usepackage{algorithm}
\usepackage{algorithmic}
\usepackage{multirow}
\usepackage{flafter}

\begin{document}
\title{When Machine Learning Meets Big Data: \\A Wireless Communication Perspective}


\author{
\IEEEauthorblockN{Yuanwei Liu,~\IEEEmembership{Senior Member,~IEEE,}
Suzhi Bi,~\IEEEmembership{Senior Member,~IEEE,}
 Zhiyuan Shi,~\IEEEmembership{Member,~IEEE,}
 and Lajos Hanzo~\IEEEmembership{Fellow,~IEEE}
 }

\thanks{Y. Liu is with Queen Mary University of London, London, U.K., email: yuanwei.liu@qmul.ac.uk.}
 \thanks{S. Bi is with Shenzhen University, Shenzhen, China, email: bsz@szu.edu.cn.}
\thanks{Z. Shi  is with Onfido, London, U.K., e-mail: zhiyuan.shi@onfido.com.}
\thanks{L. Hanzo  is with University of Southampton, Southampton, U.K., e-mail: lh@ecs.soton.ac.uk.}
}

\maketitle

\begin{abstract}
We have witnessed an exponential growth in commercial data services,
which has lead to the 'big data era'. Machine learning, as one of the
most promising artificial intelligence tools of analyzing the deluge
of data, has been invoked in many research areas both in academia and
industry.  The aim of this article is twin-fold. Firstly, we briefly
review big data analysis and machine learning, along with their
potential applications in next-generation wireless networks. The
second goal is to invoke big data analysis to predict the requirements
of mobile users and to exploit it for improving the performance of
``social network-aware wireless". More particularly, a unified big
data aided machine learning framework is proposed, which consists of
feature extraction, data modeling and prediction/online
refinement. The main benefits of the proposed framework are that by
relying on big data which reflects both the spectral and other
challenging requirements of the users, we can refine the motivation,
problem formulations and methodology of powerful machine learning
algorithms in the context of wireless networks.  In order to
characterize the efficiency of the proposed framework, a pair of
intelligent practical applications are provided as case studies: 1) To
predict the positioning of drone-mounted areal base stations (BSs)
according to the specific tele-traffic requirements by gleaning
valuable data from social networks. 2) To predict the content caching
requirements of BSs according to the users' preferences by mining
data from social networks. Finally, open research opportunities are
  identified for motivating future investigations.
\end{abstract}


\section{Introduction}


The family of next-generation techniques appearing on the horizon aim
for providing high-quality communication services relying on a high
throughput, massive connectivity and low delay. For cellular systems,
$1$-$10$ Gbps downlink transmission rate and less than $1$ millisecond
delay is expected. Meanwhile, 5G standards also allow conventional
high-speed cellular communication to coexist to be with
Machine-to-Machine (M2M) and the Internet of Things (IoT) services
with an emphasize on large coverage, advanced dense connectivity
(associated with up to one million sensor connections within a square
kilometer area) \cite{Bi2015Bigdata}. For instance, IoT networks
conceived for industrial control and health monitoring generate vast
amounts of sensing data. In addition, the autonomous connected
vehicles of the near future will support millions of high-velocity
devices, which would significantly increase the data traffic of the
emerging next-generation (NG) network.

\begin{figure}[t!]
    \begin{center}
        \includegraphics[width=3.8in]{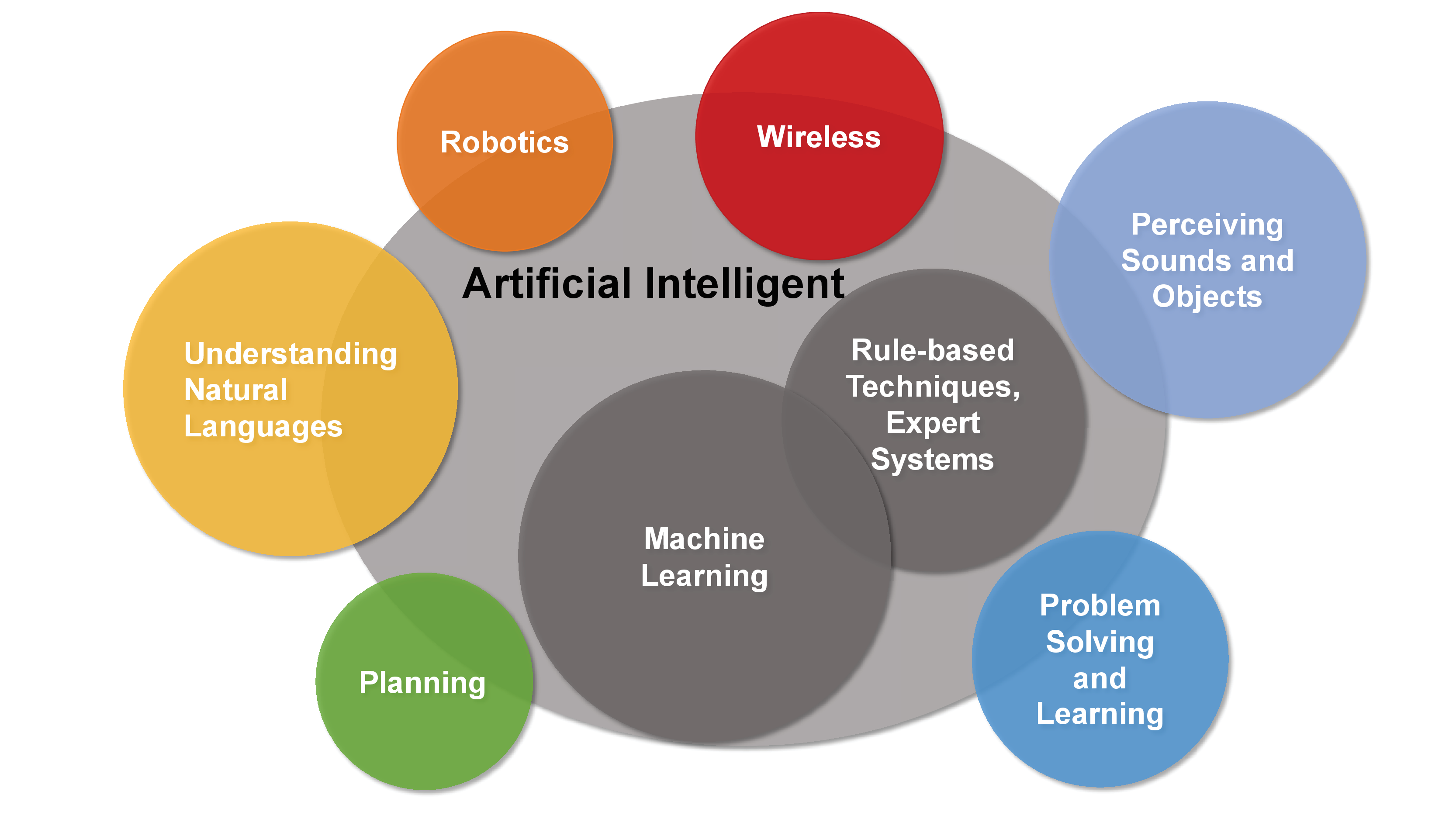}
        \caption{Relationships between artificial intelligence and machine learning.
In a rule-based system, the static knowledge is explicitly represented as a set of rules that can be thought of as a collection of facts, constraints or regulators. An expert system constitutes one of the classic examples of a rule-based system that makes use of various domain-specific expert programs. The implementation and integration of heuristic knowledge also provides a basis for transparency and flexibility. By contrast,  machine learning, tends to automatically learn computational models based on its access to huge amounts of data without relying on hard-coded routines of specific instructions.}
        \label{Relationships}
    \end{center}
\end{figure}

The term  'Artificial Intelligence' (AI) was first coined by John McCarthy dating back to 1956 \cite{McCarthy2007ai}. As a branch of computer science, AI promises to enhance the intelligence of computers by imitating the actions of  human beings, i.e., understanding natural language, planning, perceiving sounds and objects as well as in problem solving and learning, as illustrated in Fig. \ref{Relationships}. There are numerous techniques of formulating AI solutions. The early seminal approaches tend to explicitly program a decision system by leveraging domain-specific knowledge, leading to the concept of expert systems. Various carefully defined rules have to be structured and shaped for creating such domain-specific expert programs. In contrast to the expert systems containing millions of lines of codes along with decision trees and complex rules, machine learning is of potentially lower complexity, hence it has made tremendous progress over the last decades \cite{Kibria2017Bigdata}. The stylized relationship between machine learning and AI is illustrated in Fig. \ref{Relationships}. The core motivation behind machine learning is that of allowing automonous learning/training based on its  access to huge amounts of data, rather than writing hard-coded routines of specific instructions. The resultant algorithm then offers a variety of intelligent actions, ranging from learning based on past experience, reasoning for comprehending complex ideas and generalizing to new situations.

In this light, a natural question arises: how can big data and machine learning help to enhance the performance of future 5G and beyond wireless networks?  At the time of writing it is not so widely understood how to harness machine learning techniques for solving the typical optimization problems in wireless networks scenarios relying on big data analysis. Contributing to the solution of this problem motivates us to develop this treatise, where the big data resources are exploited by analytical machine learning tools, in support of intelligent applications in wireless networks. The main contributions of this article can be summarized as follows. 1) The beneficial exploitation of big data resources in wireless networks are discussed; 2) Three classification approaches suitable for machine learning are proposed according to  different criteria. Both the pros and cons are discussed by considering compelling application scenarios. 3) A unified framework of invoking machine learning techniques in social network-aware wireless is proposed and augmented by considering a pair of intelligent application scenarios.

The rest of this article is organized as follows. In Section II, we introduce the forthcoming  `big data era', followed by the classification and the pros as well as cons of machine learning in Section III. Then we propose a unified big data aided machine learning framework in Section
IV, followed by discussing a pair of promising case studies in Section V.  Finally, our conclusions and future research scenarios are discussed in Section VI.


\section{Big Data in Wireless Networks}

In this section, we will classify the family of existing data sources into three broad categories, namely general wireless data, social network-aware data (social data) and cloud data, as illustrated in  Fig.~\ref{fig:bigdata}. The corresponding application scenarios are discussed below.

\subsection{Wireless Data}
\begin{figure*}[htbp]
    \begin{center}
\includegraphics[width=0.9\textwidth]{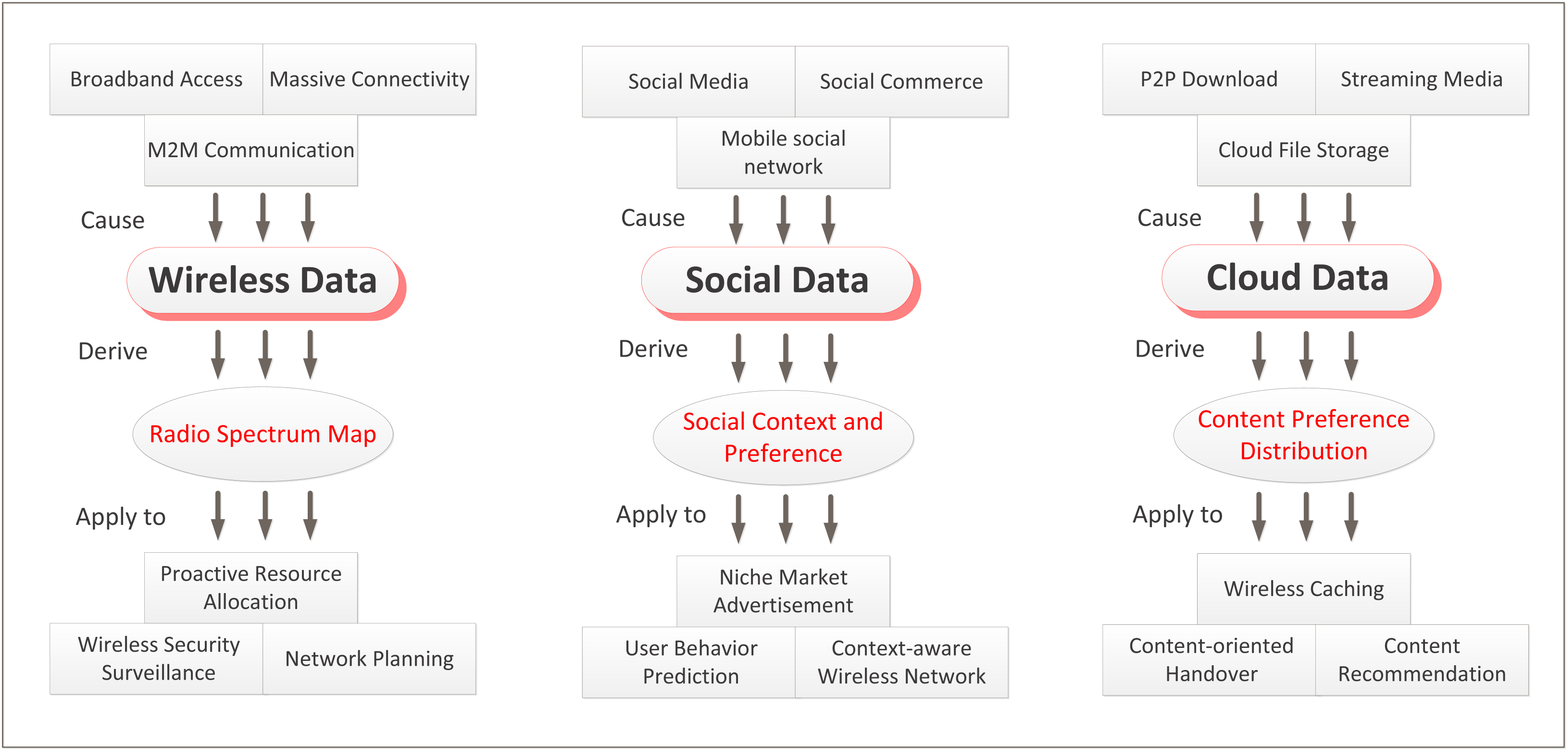}
\caption{Wireless big data: its source, the information hidden in it and its applications. }
\label{fig:bigdata}
    \end{center}
\end{figure*}

The big data generated by wireless users contains useful information about their activity patterns versus time, frequency and space. For instance, we can infer from the data traffic/demand variation over time, the interference power at different frequencies, and the congestion level distribution at different locations, etc. By exploiting these spectral patterns, we can efficiently manage the wireless resources for improving the system's spectral efficiency and for enhancing the users' service quality. As shown in Fig.~\ref{fig:bigdata}, one of these intelligent applications is load balancing relying on proactive wireless resource allocation. In this context the operator can adjust the transmit power, frequency or direction (e.g., through sectorized antennas) of different base station transmitters relying on the estimation of the mobile users' distributions. Furthermore, the operator can dispatch mobile base stations in advance when anticipating a regional surge of data traffic. Another important application is constituted by  wireless security surveillance. Given the spectral activity patterns inferred we can detect anomalies in the radio environment, such as the perspective of rouge base stations, based on atypical measured real-time spectrum usage. The key challenge of the above applications is to derive such a ``radio map" from the vast amount of noisy wireless big data, so that we can accurately characterize the spectral usage patterns in different dimensions and scales \cite{Bi2017Radiomap}.

\subsection{Social Data}

The main cause of the soaring data volume in the Internet is online social networks. The penetration of the mobile Internet into our daily lives  makes convenient multimedia communications ubiquitously accessible for everyone. The amount of social data has reached to an astonishing magnitude and it is set to maintain an increasing tendency in the next few years. In 2017, on average over $500$ million Tweeter messages per day were generated, and more than $80\%$ of them were initiated from mobile terminals (data from online). Social network data on one hand is featured by its strong ties to public events in the physical world. For instance, an important football match or political event may inspire heated online discussions that may last for days; meanwhile, frequent sharing of high-score online evaluations about a movie premiere or a newly opened restaurant may attract a large number of customers in the real world. On the other hand, mobile social network data contains rich information about the contexts/preferences of individuals or social groups. For instance, we can infer from the tweets that a group of mobile users are at a famous tourist site, but they are unhappy about the wireless services. Accordingly, we can improve the typical tourist-experience by temporarily allocating more bandwidth to the nearby base stations. As a result, a social network-aware wireless concept can be conceived. Here, the major technical difficulty is to infer the true ``meaning" of the users' messages, based on which appropriate  actions can be taken. Given the vast amount of social network data and the diverse nature of information conveyed, e.g., text or multimedia, this may be achievable through advanced machine learning.

\subsection{Cloud Data}

Another major cause of data tsunami in wireless network is the transmission of multimedia contents stored in the cloud servers. It is predicted that by 2019, over $80\%$ of the world's Internet traffic will be videos, such as Youtube short clips, Netflix long clips, and Facebook live streaming (data from online). Meanwhile, online audio streaming services also contribute a large fraction of the remaining $20\%$ data traffic. A unique feature of cloud-based big data is that the users' preferences concerning a certain content are often similar and correlated. For instance, $3\%$ of YouTube videos accounts for over $90\%$ of its total views (data from online). In other words, most contents transferred over the Internet are based on its popularity. We can therefore reduce the tele-traffic of the system by exploiting the users' preference for specific cloud contents. For instance, we can pre-cache the most popular videos at the edge servers, so that no real-time backhaul data downloading is needed for frequent requests. Besides, by caching the video contents at multiple nearby base stations, we can form a virtual antenna array and support seamless handovers for mobile terminals. Additionally, each individual user's preferences for specific multimedia contents, if available, can be used for predicting the user's future demand, based on which the network operator can perform pre-feeding or recommendation actions. Naturally, the main technical challenge is to accurately predict the users' preference distribution. Another challenge is to correctly label the tremendous amount of videos for future reference/searching. In this case, using manual labeling is infeasible for high-population networks, hence artificial intelligence techniques are needed for designing content-oriented intelligent wireless transmissions.

\section{Machine learning in Wireless Networks}

\begin{figure*}[htbp]
    \begin{center}
        \includegraphics[width=6.8in]{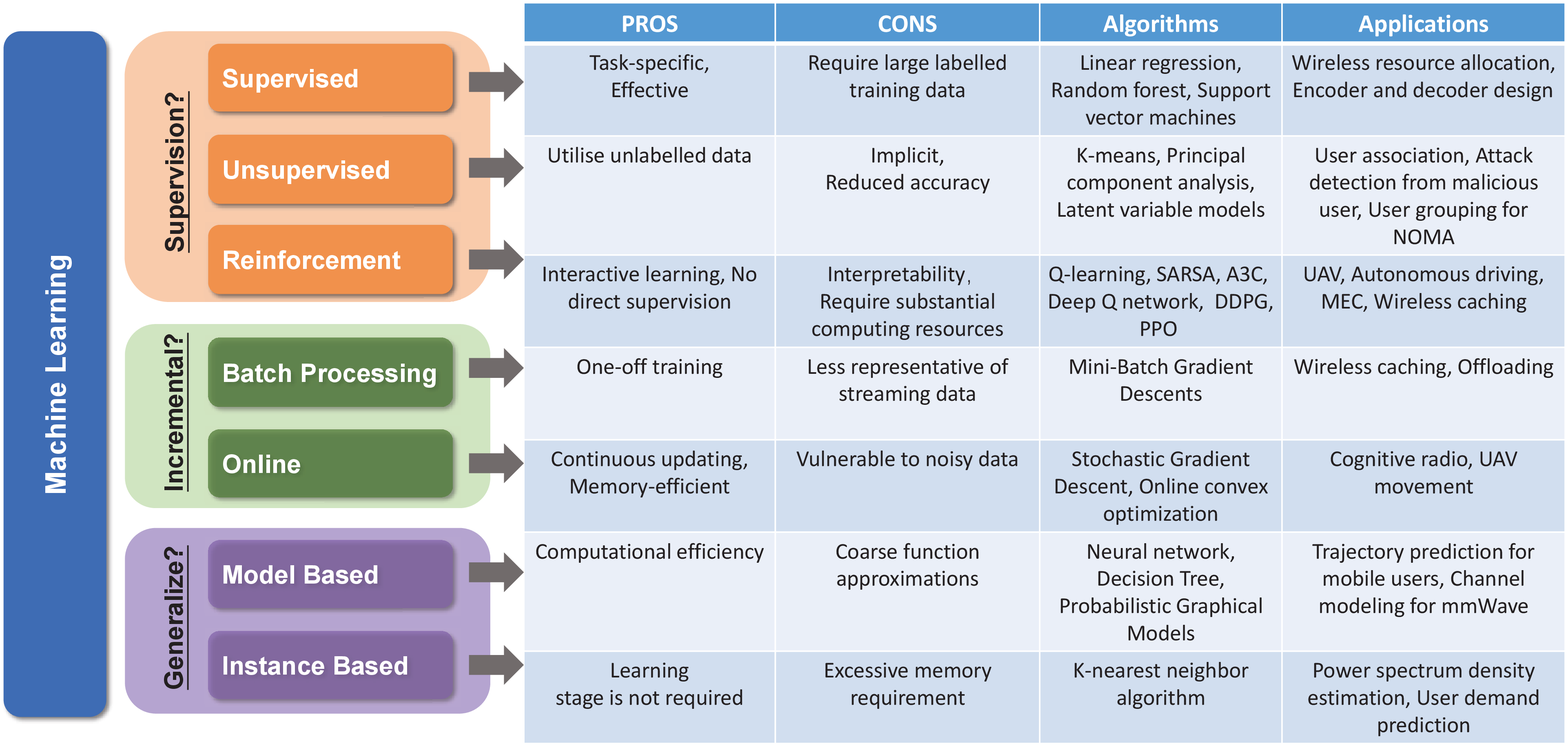}
        \caption{Classification of machine learning, the corresponding pros/cons and application scenarios. The above table aims to provide a systematic review of the practical applications of modern machine learning methods in wireless communication. By categorising the various approaches into three principal groups according to 1) how they require human supervision 2) are they able to perform incremental learning 3) how they generalise large scale training/testing scenarios, we explicitly demonstrate the pros and cons of each scheme and highlight the most appropriate applications.}
        \label{machine learning}
    \end{center}
\end{figure*}

 In this section, we will detail how we classify the machine learning techniques and how they can be applied to wireless communications. Moreover, the pros and cons of each machine learning family are discussed, as illustrated in Fig. \ref{machine learning}.


\subsection{Human Supervision Requirement}

According to whether or not the algorithms require human supervision, the machine learning models may  fall into three general categories \cite{Nasrabadi2007ml,Jiang2017ML}.
\begin{itemize}

\item {\bf Supervised Learning} is considered to be a major branch in machine learning, which has been extensively studied and developed. Large quantities of human-labeled data should always be readily available for learning a functional mapping between the training samples observed and the desired output. The advantage of supervised learning is that both the convergence speed and the action quality are high, although they typically require a large amount of data to be labeled manually, which makes the data processing more complex. Attractive scenarios for applying supervised learning can be wireless resource allocation, encoder and decoder design, where the objective function definition of the application is clear and collecting sufficient training data is relatively easy and less costly.

\item {\bf Unsupervised Learning} relies on vast amounts of unlabeled data for inferring the underlying information structure,  without depending on external resources and human supervision. The advantage of unsupervised learning is that no prior knowledge is required. However, this comes at the cost of potentially reducing its accuracy. Another disadvantage of unsupervised learning is that the automatically discovered data  is not always representative of real world conditions. Given the unique features mentioned above, unsupervised learning is suitable for solving the problems of user association, user grouping for hybrid multiple access, attack detection for malicious users, etc.

%
%

\item {\bf Reinforcement learning} was initially designed to discover optimal action spaces through adaption and interactions in uncertain time-varying environments, which provides another way of learning from unlabeled data, as long as either positive or negative feedback can be gleaned during the process of learning by trial and error. In contrast to supervised/unsupervised learning, reinforcement learning does not require direct supervision. Moreover, the interactive learning paradigm has the capability of learning to act, so as to adapt itself for achieving an ever-improving performance. The disadvantage of reinforcement learning is however that it relies on huge amounts of resources. Another drawback is that the resultant high-performance solutions often lack plausible physical interpretations. Nonetheless, successful application scenarios for reinforcement learning have been found in unmanned aerial vehicle (UAV) communications, autonomous driving, mobile edge computing and wireless caching placement, just to name a few.

%

\end{itemize}

\subsection{Learning Capability}

Based on their learning capability, we can classify the family of machine learning models into two subsets \cite{Anderson2009online}.
\begin{itemize}

\item {\bf Batch learning } based algorithms typically train a model after a sufficiently high amount of training data has been collected prior to the learning task. This offline learning manner has the advantage of allowing for more involved machine learning based on all the available data, where the model tends to be updated infrequently in general. The main advantage of batch learning is that the convergence speed is high. However, batch learning may not be suitable for real-time learning of rapidly fluctuating processes subject to stringent delay requirements. As a result, beneficial application scenarios can be wireless caching or wireless offloading.
%

\item {\bf Online learning } enables a model to learn from a stream of data instances arriving sequentially, which is achieved by continuously changing and adapting its structure and parameters. The `on-the-fly' learning scheme holds the promise of being both memory efficient and highly scalable in solving large-scale learning problems. A particularly remarkable advantage of online learning is that it is eminently suitable for real-time processing. Nevertheless, its convergence speed typically requires slow. Therefore, online learning can be used for cognitive radio networks or UAV movement scheduling.

\end{itemize}

\subsection{Generalization Requirement}

On the basis of how the methods generalize their findings from the training data to hitherto unseen examples, the machine learning models can be split into two parts \cite{robert2014machine}.
\begin{itemize}

\item  {\bf Instance-based learning} algorithms  \cite{Daelemans2005Cambridge} tend to use the whole set of instances found in the training data stream for constructing inference structures and for predicting unseen instances. Although they are quite capable of competent generalization, this is achieved at the cost of extended search time and high memory requirements. An especially remarkable advantage of instance-based learning is that it does not require any prior-model assumptions. Therefore, instance-based learning is suitable for complex wireless scenarios, such as power spectrum density estimation or user demand prediction. However, large data sets are required for high-quality instance-based learning.

%

\item {\bf Model-based learning}  aims for finding the optimal parameters of the algorithms designed \cite{Quinlan1993combining}, so as to optimize the objective function and to maximize the generalization capability in the face of hitherto unseen testing data. Such models usually suffer from limited accuracy,  but offer higher computational efficiency. Hence the key advantage of model-based learning is that its implementation cost is low. Given these attributes, model-based learning can be readily used for the movement trajectory prediction of mobile users, for channel modeling estimation in millimeter wave communications, etc.

%

\end{itemize}

\section{A Unified Big Data Aided AI Framework}

\begin{figure*}[htbp]
    \begin{center}
        \includegraphics[width=5.8in]{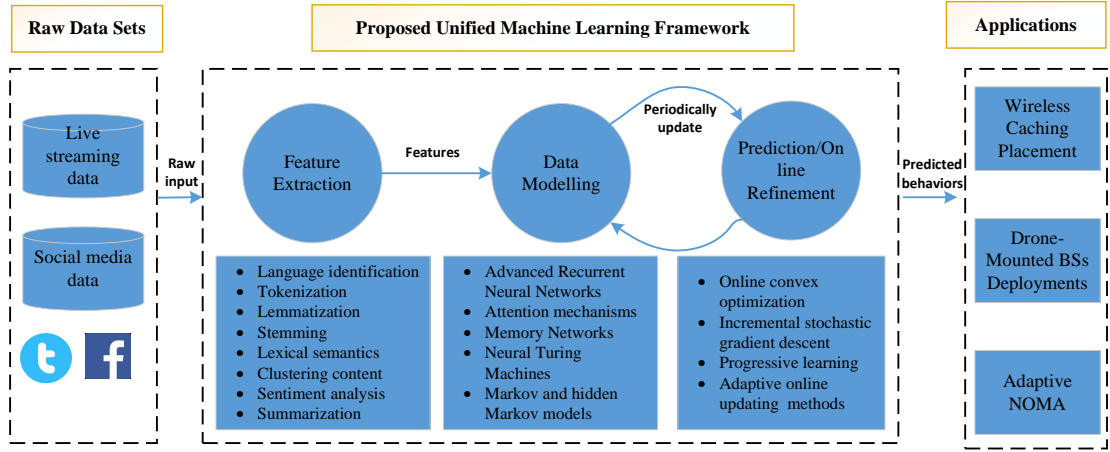}
        \caption{A Unified Big Data Aided Machine Learning Framework.}
        \label{Unified framework}
    \end{center}
\end{figure*}

In this section, we propose the new unified machine learning framework in  Fig. \ref{Unified framework}. According to social data, cloud data or wireless data, we aim for invoking advanced machine learning approaches for analysing the data for extracting useful information. More particularly, the proposed framework includes three stages, as illustrated in Fig. \ref{Unified framework}, which are detailed in the following.

1) \textbf{Feature Extraction}: during this stage, we aim for extracting feature from the data sets, which contain the geo-location tags of mobile users. Possible data sets can be social, wireless data or cloud data, as illustrated in Fig. \ref{fig:bigdata}. Let us consider social data as an example,  including social media, social commerce or mobile social networks. We can derive social context and user preferences with the aid of social data by invoking natural language processing or other techniques. The detailed procedure includes a pair of steps. The first step is  syntax processing: the input is the tremendous amount of social data, which is subjected to natural language processing techniques, such as language identification, tokenization, lemmatization, stemming, etc. The output should be some useful geo-and-time-tagged  key words according to the specific application, say crime prevention. The second step is semantic analysis: The input for this step is constituted by the key words subtracted during the last stage, with the aid of applying sophisticated  techniques, such as lexical semantics, clustering content, sentiment analysis, and summarization. The outputs are expected to be context-aware user preferences. A suitable example is to apply the Twitter data to predict the hot topics of high popularity for content caching in wireless networks. 

2) \textbf{Data Modelling}: the feature extraction stage mentioned before can be regarded as the pre-data processing. This data modelling stage constitutes the core process of establishing a machine learning model. More particularly, the feature we have extracted from the last stage such as the user mobility and content popularity of users represent the data set we seek for modelling the problem.  Usually, we first formulate an objective function (e.g., maximizing the QoE, reducing delay or improving energy efficiency) for enhancing the performance of  wireless networks considered. Then we invoke the modelling strategies listed in the data modelling stage of Fig. \ref{Unified framework} for mathematically representing the formulated problems.

3) \textbf{Prediction/Online Refinement}: After establishing the machine learning model, we proceed either to the prediction or to the online refinement stage. For the prediction, this implies that the model established  can  only be used for predicting the user behavior/network performance without updating the model invoked. Regarding the online refinement, this means that the model established can be refined based on changes in the real-time data input. The intuitive difference between prediction and online refinement is that the system provides a flexible mechanism for inference testing either relying on an off-line model or on a periodically refined model. The selection of these two variants ultimately depends on the particular context of a specific application. For the use case when the data distribution does not show large variations over the pertinent time scales, prediction relying on an off-line trained model is always the best choice for ease of computation. On the other hand, online refinement has to be the undertaken for improving local errors, when the data are intrinsically time-varying. However, the latter is  computationally more demanding. Note that a remarkable advantage of this framework is that the machine learning model depends not only on the historical data, but also on real-time data. For example, we could periodically use the past two weeks of data to predict the next day's trends. By exploiting these time sequence characteristics, the prediction capability of this model becomes quite accurate and timely. The techniques listed in the prediction/online refinement stage of Fig. \ref{Unified framework} can all be beneficially applied.



\section{Case Study: Social Network-Aware Wireless}

Having introduced the three key stages of the proposed framework, the next important step is to identify suitable application scenarios. In this section, we will consider a pair of intelligent applications for demonstrating how to apply our big data aided machine learning framework for fostering intelligent wireless networks. In both case studies, we exploit the position information collected by using \emph{Twitter's API} to infer the periodic behaviors of users in social networks. Efficient algorithms based on neural networks are proposed for predicting the user's mobility or content popularity, which belongs to the feature extraction stage. Then reinforcement learning is invoked for solving the problems formulated  in the data modeling stage, either for cooperative caching or UAV deployment, for example. 

\subsection{Dynamic UAV Deployment and Movement Design}

Thanks to the rapid capability-upgrade of drones, base stations (BSs) on-the-fly have emerged as an efficient solution for improving the radio coverage conditions in the face of rapidly fluctuating traffic demands by appropriately adjusting the drone positions.
Given the availability of drone-mounted BSs, how to position them for satisfying the dynamically evolving data requests and hence maximizing the benefits to operators becomes one of the most
challenging and critical problems. In such a scenario, the strategy to deploy the drone-mounted BSs can be categorized into the following two types:
a) Pre-deploy UAVs before the requests come. Based on the proposed framework, the historical social data can be exploited for predicting the forthcoming data requests in a certain area. If the outputs of data mining indicate that data requests may be expected to exceed the capability of fixed BSs covering that area, drone-mounted BSs can be deployed to that area in advance. The benefit of this strategy is to provide improved user experience by enhancing the network's capability before congestions happen. b) Move UAVs based on real-time requests. Naturally, historical social data cannot exactly predict the user requests since they fluctuate dynamically. Therefore, a real-time UAV-action is desired for improving the network's capability to improve user experience. The real-time social data related to the area of interest, such as tweets complaining about poor network quality, can be further analyzed and more drone-mounted BSs can be dispatched to the hot-spot area. More particularly, by analyzing the complaining tweets, we first decide the satisfaction level of users about the current transmit rate based on the quality-of-service (QoE) model, which identifies the potential tele-traffic congestion events. Finally, the required number of UAVs, as well as the movement of UAVs, can be decided for maximizing the sum mean of score (MOS) of users by adopting machine learning approaches.

\begin{figure*}[t!]
    \centering
    \subfigure[Procedure for UAV deployment and movement]{\includegraphics[width= 3.1in, height=2.1in]{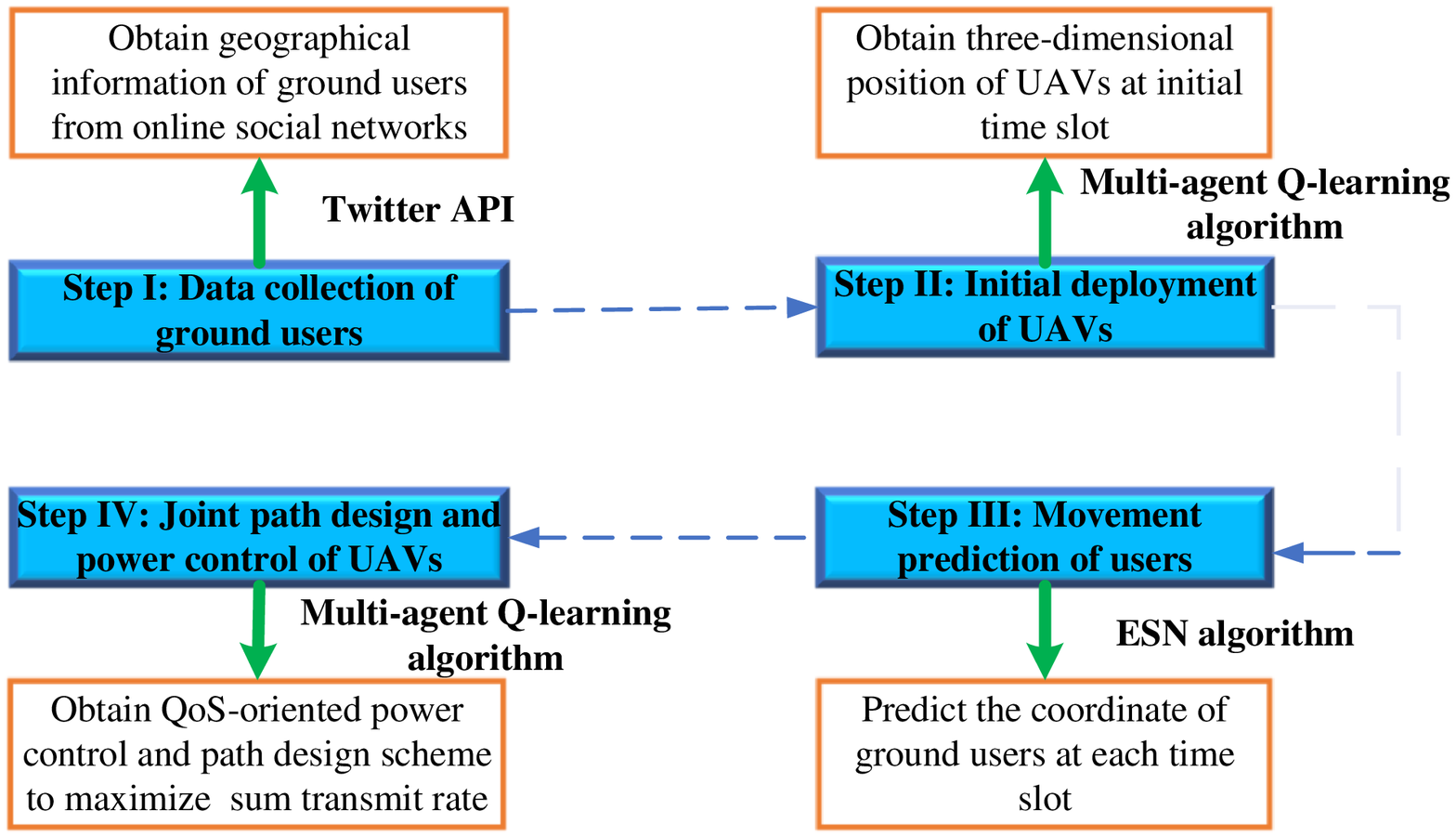}}
    \subfigure[Trajectory for UAV deployment and movement]{\includegraphics[width= 3.1in, height=2.1in]{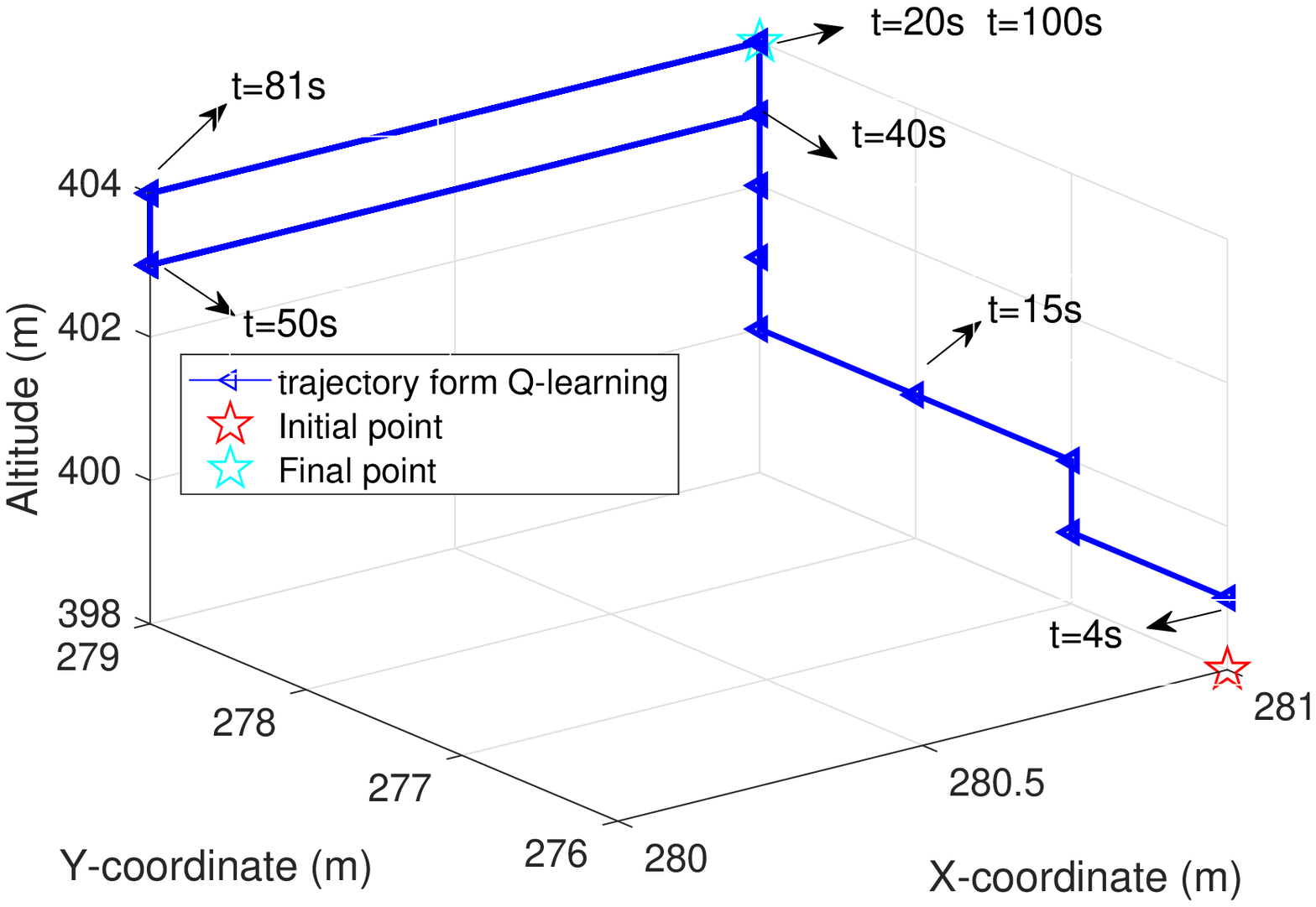}}
    \caption{Case study of deployment and movement Design for multiple-UAV networks: ~\cite{Xiao2018UAV}.}
    \label{Case study for UAV}
\end{figure*}
In this case study, multiple UAVs are used as flying BSs for serving ground users by  considering both of the aforementioned two types of UAV deployments. Each UAV is capable of roaming in a free 3D space. The ground users are supposed to move flexibly. Therefore, UAVs can fly dynamically according to the real-time position information of users. Fig. \ref{Case study for UAV} (a) illustrates the four steps of UAV deployment and movement design relying on the proposed framework. Step I can be regarded as the \emph{feature extraction} stage. Here the movement of users is the key feature we would like to extract. Regarding the \emph{data modelling} stage, multi-agent Q-learning is the core algorithm invoked at this stage, which is employed  both in Step II and Step IV for designing the specific deployment and movement of UAVs. Fig. \ref{Case study for UAV} (b) illustrates how a UAV moves from an initial position to its final destination with the aid of applying reinforcement learning for formulating its trajectory. In our scenarios, multiple UAVs move simultaneously  according to the prediction of the movement of ground users, which can be regarded as the \emph{prediction stage}. By doing so, the pre-deployment of UAVs can be achieved for enhancing the energy efficiency of networks.

\begin{figure}
\centering
\includegraphics[width=3in]{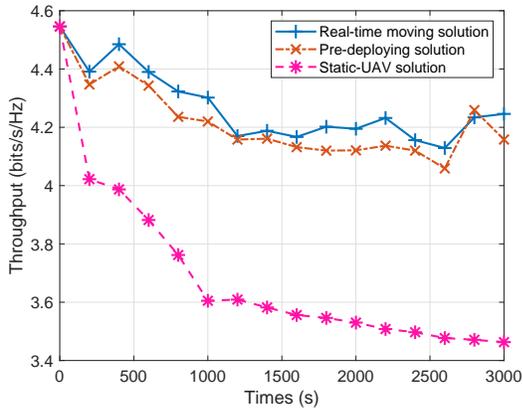}
\caption{Comparison of throughput between moving-UAV solution and static-UAV solution.}\label{throughput}
\end{figure}

Fig. \ref{throughput} characterizes the comparison of the throughput between the moving-UAV solution (both pre-deployed and real-time movement solution) and static-UAV solution. It can be observed that the instantaneous transmit rate decreases, as time elapses. This is because the users are roaming during each time slot, and when the density of users is reduced, the instantaneous sum of the transmit rate is affected. It can also be observed that re-deploying UAVs based on the movement of users is an efficient method of mitigating the downward trend compared to the static scenario.

\subsection{Wireless Caching Placement and Resource Allocation}

Due to the rapid development of mass-storage techniques, storage becomes an increasingly low cost resource, whilst the opposite trend prevails for spectral resources. The caching process is constituted by two parts: caching placement and cached content delivery. In contrast to most of the existing research contributions, which have assumed that the caching placement process follows a specified distribution, in this case study, the content caching at each BS is dynamically varied according to the requirements of users. More particularly, with the aid of social data within a given area and the corresponding location information of BSs, the caching placement can be beneficially improved. The main objective of caching placement is typically to identify the most popular $K$ files to be stored in the BS during a particular period in order to reduce the potential latency to download these files by the users. To efficiently allocate wireless contents, such as video clips to base stations, some prior information is required, on user mobility and content popularity. However, the complicated relationships between historical information and future information make conventional approaches less applicable. AI algorithms are capable of predicting the demands of the networks in the near future, which helps us to efficiently cache most popular contents when the communication resources (e.g., bandwidth, storage capacity and computing speed, etc) are finite. Considering some of the recent vile terrorist attacks as an example, typically there are immediate reactions in social media, such as Facebook and Twitter. In the ensuing period the discussions, related posts and messages on the social data will increase. By invoking natural language processing to analyse the posts, the key words can be extracted. Then the key words may be classified or mapped to different hot topics. The different hot topics here decide what contents should be cached in particular base stations. Based on the time-geo-tagged hot topics, machine learning may be invoked for training this learning model. The final goal is that according to the historical key words in a certain past period of say 24 hours, the content to be cached in the next period of 2 hours for example can be dynamically decided.

\begin{figure*}[t!]
    \centering
    \subfigure[Flowchart for Wireless Caching]{\includegraphics[width= 3.1in, height=2.1in]{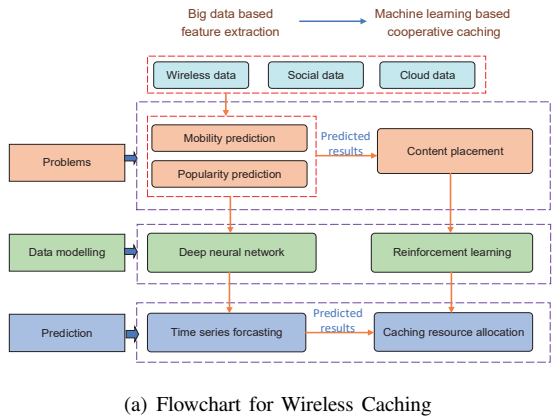}}
    \subfigure[Simulation Results for Wireless Caching.]{\includegraphics[width= 3.1in, height=2.1in]{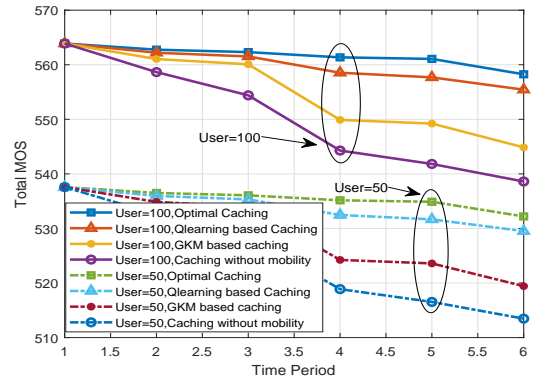}}
    \caption{Case study of content placement in wireless cooperative caching  \cite{Yang2018cachinng}.}
    \label{Case study for wireless caching}
\end{figure*}

Fig. \ref{Case study for wireless caching} (a) illustrates the detailed procedure of wireless caching. More particularly, we can still map the wireless caching placement to our proposed framework. In the feature extraction stage, both the mobility of the users and the content popularity have to be predicted for supporting proactive content placement. During the data modelling stage, neutral networks and reinforcement learning may be invoked for mathematically modelling the mobility/content popularity and for cooperative caching placement, respectively. During the prediction stage, the output of neural networks \cite{Chen2017tutorial} can be used for predicting the cooperative caching allocation. Fig. \ref{Case study for wireless caching} (b) shows total mean opinion score (MOS) of the users in the network versus time for different algorithms. In the initial time period, the optimal contents are placed according to the users' positions. Then, since we consider a user movement scenario, the positions of users change with the time period. Therefore, the content placement is not optimal.  Hence the total MOS of users decreases. We can observe that the Q-learning algorithm, which is a close relative of the reinforcement learning algorithm, is capable of achieving a near-optimal performance, whilst outperforming global $K$-means (GKM) based caching.


\section{Future Challenges and Concluding Remarks}

In this article, the design challenges of invoking machine learning techniques for enhancing wireless networks with the aid of big data solutions have been
investigated. The key features of big data have been identified first, which spawn new research opportunities
for wireless networks. Then the classification of machine learning techniques and their corresponding application scenarios in wireless networks were discussed. Furthermore, a unified big data aided machine learning framework has been proposed. This is
followed by introducing a pair of case studies in wireless networks, namely UAV deployment/trajectory design and wireless caching placement. Yet numerous open research opportunities have to be pursued in the context of machine learning aided wireless communications, which are highlighted as follows:

\begin{itemize}
  \item \textbf{Adaptive Non-Orthogonal Massive Multiple Access}: Non-orthogonal multiple access (NOMA) constitutes a multi-faceted  technique of enhancing wireless networks. Note that each form has its own advantages and disadvantages. Therefore, different NOMA techniques are suitable for different scenarios, as illustrated in Table 9 of \cite{Liu2017NOMA}. Motivated by this, a software-defined NOMA architecture  is proposed for supporting diverse user scenarios. However, the tele-traffic  demands can vary seasonally or due to the public events. Therefore, machine learning can be invoked to predict the data traffic. The MA settings can be categorized into two types, pre-settings and real-time settings. Batch learning can be used for pre-settings. While the real-time settings are typically based on the users' feedback on social media for example to dynamically change the MA settings  on demand, which may invoke online learning, as mentioned in Section III.
    \item \textbf{Autonomous Driving in V2X networks}: Autonomous driving has the potential to benefit society in numerous ways, such as reducing traffic congestion and mitigating the environmental footprint of modern day traffic. The combination of autonomous driving and vehicle-to-everything (V2I) communications enable automated vehicles to receive up-to-date information about the neighbouring vehicles' dynamics and other traffic information, thereby enhancing both safety and traffic efficiency. The deep reinforcement learning models mentioned in Section III that are  trained by interacting with the environment, may be utilized for optimizing the behaviours of vehicles by exploring the environment in an iterative manner and learn from mistakes. With the aid of  cloud data or wireless data, the requirements of V2I networks can be updated in real-time. In this case, the autonomous vehicle becomes capable of safely reaching its destination while avoiding traffic jams with the aid of up-to-date traffic information.
    \item \textbf{Intelligent Computation Offloading}:  Mobile-edge computing (MEC) is a promising technique of meeting the ever-increasing computational demands of mobile applications by providing computing capabilities at the edge of wireless networks, while migrating the computationally-intensive tasks to the MEC server. The core concept of MEC is to provide abundant computing capabilities at the edges of networks to mitigate both the backhaul and fronthaul load and to reduce the energy consumption of mobile users. Task offloading decisions and computational resource allocation constitute a pair of challenges in MEC. Obtaining an optimal offloading policy in such a dynamic MEC system in real time is challenging. Machine learning is capable of intelligent inferences from historic information. By using the proposed framework, a real-time dynamic MEC system operating with the aid of social data can be designed by utilizing machine learning algorithms to attain significant performance versus complexity benefits.
\end{itemize}


 \end{document}